\documentstyle[preprint,pre,aps]{revtex}

\newcommand{\Vec}[1]{\bf{#1}}
\newcommand{\Mat}[1]{\widehat{#1}}

\begin{document}

\begin{titlepage}

\title{How Accurate Must Potentials Be for \\ 
Successful Modeling
of Protein Folding?}
\author{Vijay S. Pande, Alexander Yu. Grosberg\dag, 
Toyoichi Tanaka}
\address{Department of Physics and Center for 
Materials Science and 
Engineering, \\
Massachusetts Institute of Technology, Cambridge, 
Massachusetts 02139, 
USA \\
\dag {\em On leave from:\/} Institute of Chemical Physics, \\
Russian Academy of Sciences, Moscow 117977, Russia}
\date{\today}

\maketitle

\begin{abstract}
Protein sequences are believed to have been selected to
provide the stability of, and reliable renaturation to, an
encoded unique spatial fold. In  recently proposed theoretical
schemes, this selection is modeled as ``minimal frustration,''
or ``optimal energy'' of the desirable target conformation
over all possible sequences, such that the ``design'' of the
sequence is governed by the interactions between monomers.
With replica mean field theory, we examine the possibility to
reconstruct the renaturation, or freezing transition, of the
``designed'' heteropolymer given the inevitable errors in the
determination of interaction energies, that is, the difference
between sets (matrices) of interactions governing chain design
and conformations, respectively. We find that the possibility
of folding to the designed conformation is controlled by the
correlations of the elements of the design and renaturation
interaction matrices; unlike random heteropolymers, the ground
state of designed heteropolymers is sufficiently stable, such
that even a  substantial  error in the interaction energy
should still yield correct renaturation.
\end{abstract}

To appear in {\em Journal of Chemical Physics}, December 1995

\end{titlepage}

\section{Introduction}

\subsection{What is this work about?}

The native state of a protein is in a sense ``written'' in the
sequence using the ``language'' of physical interactions
between monomers. In this work, we examine the effects of
``misunderstandings'' and ``misspellings'' of this language.  

A somewhat related question was recently discussed by
Bryngelson \cite{Bryngelson}. He  considered heteropolymer
chains with {\it random} sequence and estimated the
probability that its lowest energy conformation will be
correctly detailed by the model with noisy distorted
potentials of volume interactions between monomers. The result
is that the probability, $p$, diminishes with noise amplitude,
$\eta$, as $p \sim 1 - {\rm const} \cdot \eta N^{1/2} $; for
sufficiently long chain, or in thermodynamic limit, there is
no chance to compute equilibrium conformation given that some
mistakes in the determination of energies are inevitable. 

By contrast, we consider here heteropolymer chains with
sequences that are {\it not random}, but rather ``designed''
\cite{ShakDesign}, or ``imprinted''
\cite{LongImprint}, or ``selected''
\cite{Dill}, or, in other words, obey the so-called principle
of minimal frustration
\cite{Wolynes}. We show, that for these chains the situation
is dramatically different, and there is finite probability of
successful recovery of thermodynamically stable conformation,
even in thermodynamic limit ($N \rightarrow \infty$) and for
finite non-vanishing $\eta$. 

As the work is based on rather heavy theoretical machinery, we
begin with more general introduction.  

\subsection{Protein folding in a statistical mechanics
perspective: a brief summary} 

Protein folding is one of the great challenges of modern
biophysics. Recently, there has been much progress and insight
into protein folding from the statistical physics perspective.
This came mainly with the ideas borrowed from the physics of
disordered systems, such as spin glasses
\cite{Stein}. It is now widely believed that folding of a
protein chain can be viewed as a freezing phase transition, as
in spin glasses. The sequence of types of monomers along the
chain leads to quenched disorder, and polymer bonds between
monomers impose frustrations. The concept of a freezing
transition seems to resolve one of the mysteries surrounding
the problem, namely, why the native state of a protein is
realized via a unique conformation, at least, in a coarse
grained sense. 

One of the main differences between proteins (and biopolymers
in general) and other systems more familiar to physicists is
the way in which quenched disorder appears. While in  regular
spin glasses and similar ``un-animated'' systems, the
appearance of any particular realization of the disorder is
far from any control, in biology the situation is dramatically
different. First, the biosynthesis of proteins (as well as
other biopolymers) manages to produce {\it macroscopic amount
of identical copies} --- a situation unthinkable in other
disordered systems. Furthermore, the sequences of the existing
real proteins are believed to be the product of evolution and
thus  are of great interest. Do they have some distinct
properties compared to other possible sequences? Do they
provide  some general properties of proteins, such as their
ability to form the unique spatial fold, or only particular
conformation and/or function of each particular protein? 

To this end, it now seems  well established  that the ability
of polymer chain to undergo a freezing phase transition into
the state with a unique (or almost unique) conformation is
common to many models of random heteropolymers. In other
words, to achieve the uniqueness of the ground state
conformation one does not have to impose any requirements on
the sequence, which can be therefore chosen even at random.
This was first shown for the so-called independent-interaction
model \cite{Garel,IndepenInt} (which is in a sense even more
random than random heteropolymer: instead of $N$ independent
monomers, it has about  $\sim N^2$ independent
monomer-to-monomer interaction constants). This argument was
extended for  random copolymers with two types of monomers
\cite{Sfatos} and brought into the most general form in Ref.
\onlinecite{RandHet}.

In independent (and even earlier) development, the hypothesis
was made that the sequences of real proteins deviate from
random in such a way that to satisfy the requirement of the
so-called ``{\it minimal frustration principle}''
\cite{Wolynes}. In a sense, this idea goes back to even
earlier ideas of Abe \cite{Abe}.  Recent developments, related
mainly to computer Monte Carlo simulations of freezing
kinetics, reveal the insight into the role of minimal
frustration as a factor pulling down the energy of the ground
state conformation and thus providing the gap in the energy
spectrum necessary for reliable folding \cite{ShakDesign}.

Thus, it seems that kinetic reliability of folding requires
that sequences of proteins are not random, but ``edited.''
Statistical analysis reveals indeed, that although the
sequences are close to random \cite{Ptitsyn}, the systematic
deviations from randomness do exist \cite{ProtCor}, and they
are at least compatible with the idea of energy optimization
of the ground state.

The question appears how to model the ensemble of sequences
more realistically then by taking them at random. Two
approaches has been recently suggested. Both employ  physical
interactions between monomers to build up polymeric sequence,
and are both based on energy optimization of the native
conformation. The first approach, due to Shakhnovich and Gutin
\cite{ShakDesign}, implies the search for optimization in 
``{\it sequence}'' space by swapping the monomers along the
chain while preserving (native) conformation.  Apart from the
speculative possibility to model evolution, this is obviously
intended for computer simulation. On the other hand, we
suggested energy optimization in the ``{\it monomer soup}''
prior to polymerization \cite{LongImprint}.  Although these
two models are considerably different in spirit, they appear
to be identical from the point of view of mean-field
theoretical treatment. 

The freezing transition of an ensemble of sequences which have
been energetically optimized (``designed'') for a particular
conformation was examined for the black-and-white model (with
two types of monomers) in Refs. \onlinecite{q2Design,Sharad};
the analysis was extended for the general case in Ref.
\onlinecite{Design}.  For what follows, it is important that
three globular (that is, compact) phases were found on the
phase diagram of thus ``designed'' or ``imprinted'' polymer:
we call them random, frozen and target. The random globular
phase is pretty much the same as the globule of a
homopolymer,  it is comprised of vast number of conformations
(all of which are compact). By contrast, frozen and target
phases   are each comprised of one or very few conformations.
In the target phase, this unique conformation is exactly the
one which is targeted by the design procedure. On the other
hand, in the frozen phase, the system freezes to the
conformation which is unrelated to the design conformation
and  therefore cannot be controlled.

\subsection{Sequence design and folding are governed by {\it
different} interactions}

As  sequence design is based on energy optimization, it
employs physical interactions between monomers. It is however
possible, and, moreover, almost inevitable, that these
interactions are somewhat different from those governing
folding. Apart from speculations on the interactions that
governed  the ``design'' of modern proteins by evolution, we
mention three illustrations of our thesis:
\begin{enumerate}
\item{When one tries to find theoretically or computationally
the native state for a chain with a given sequence (direct
protein folding problem), one can say that nature details the
interactions used in the design of  protein sequences and
man-made potentials are used as substitutes in the simulations
of renaturation.} 
\item{Similarly, when one is looking for a sequence to fold
into a given conformation, one is essentially trying to design
the sequence using artificial potentials in such a way, that
this sequence under real natural interactions will fold in a
desirable way.}
\item{Speaking of the attempt to reproduce protein-like
properties in the man made heteropolymer via the Imprinting
procedure \cite{LongImprint}, we have to acknowledge some
difference between interactions of monomers in the soup prior
to polymerization and interactions of the links of polymer.}
\item{ One can consider the renaturation of a protein in a
solvent different than that used during ``design'' also as an
experiment in which the interactions during design and
renaturation are different. }
\end{enumerate}
If there are, say, $q$, different monomeric species involved
in our polymer ($q=20$ for proteins), interactions between
species $i$ and
$j$ can be described in terms of the $q
\times q$ matrix $B_{ij}$.  In general, there are two
different matrices, $B_{ij}^{p}$ and
$B_{ij}$: the first  governing preparation and second 
governing folding behavior of the already prepared chain. 

To have two {\it different}  interaction matrices for design
and renaturation is somewhat similar to writer and reader who
use different languages. {\em Naprimer, my nadeemsya, chto
nash chitatel' schitaet etot tekst  napisannnym po-angliiski i
poetomu vryadli poimet etu  frazu} \cite{Russian}. Clearly,
such a venture has a chance if and only if those languages are
not completely different, but merely dialects of one language.
Similarly, infinitessimally  small changes to the interaction
matrix should not have any significant  ramifications, while
on the other hand, a radically different matrix  structure
should lead to completely different folding behavior.  Using
the terminology of frozen and target phases, we can ask if the
chain designed with some matrix 
$B_{ij}^{p}$ will freeze to target state when governed by
another matrix
$B_{ij}$? In other words, if we want to get the target phase,
how accurate should we be in choosing matrices $B_{ij}^{p}$
and $B_{ij}$? Another interesting aspect of the question is
which properties of
$B_{ij}^{p}$ and $B_{ij}$ matrices are important, that is to
which of them the chain behavior is sensitive? And what
measure do we use to define the proximity of interaction
matrices? 

Previous treatments \cite{SGmutation,Bryngelson} have
addressed certain aspects of these questions.  However, they
differ from the present work in that we  model the effect of
evolutionary optimization and the nullification of this by
errors in the interaction potentials, whereas Refs.
\onlinecite{SGmutation,Bryngelson} examine the stability of
glassy (not evolutionarily optimized) conformations with
respect to errors.

\section{The Model}

We start from a heteropolymer chain Hamiltonian in which
interactions are described in terms of the energy of
interaction of species
\begin{equation}
{\cal H} =  \sum_{I,J}^N B_{s_I s_J} 
\delta({\Vec r}_I - {\Vec r}_J)
\label{eq:GenHam}
\end{equation}
where $B_{ij}$ is the interaction energy between monomer {\em
species} $i$ and $j$ ($i,j \in \{1 \ldots q\}$), $s_I$ is the
species of monomer at position $I$ along the chain, $N$ is the
number of monomers, and
 ${\Vec r}_I$ is the position of monomer $I$.  We use the
convention that lower case roman letters label species space,
upper case roman letters label monomer number along the chain,
and lower case greek letters label replicas.

We do not explicitly include in the Hamiltonian
(\ref{eq:GenHam}) anything leading to the overall collapse of
the chain. We do imply, however, the existence of some strong
compressing factor, such as overall homopolymeric-type poor
solvent effect (expressed with
${\cal H}^\prime = B \rho ^2 + C \rho ^3$ with species
independent $B$ and $C$ and strongly negative $B$) or box-like
external field such that the polymer is always in a globular
conformation. The particular choice of compressing factor is
known to be unimportant
\cite{Lifshits} provided that the chain is long enough; we do
not discuss here any finite size effects related to the
surface of the globule even though these might be important
for real proteins. Furthermore, we stress that this is of
vital importance for the entire approach that the chain is
maintained in the globular compact state (compare with Ref.
\onlinecite{beer}, where the design scheme failed to work just
because the requirement of overall collapsed state was
relaxed).

Since the heteropolymer sequence does not change during
folding, we immediately encounter the technical problem that
sequences are a  quenched quantity and thus we average the
free energy over all sequences (with a particular weighting
due to design)  rather than the  partition function. This
leads directly to the replica approach.  The details of the
corresponding calculation are similar to what is presented
elsewhere \cite{Design}. Here we briefly outline the main
steps. The replicated partition function can be symbolically
written as
\begin{eqnarray} &&\left< Z^{n} \right>  = \sum_{\rm sequence}
{\cal P}_{\rm sequence} 
\sum_{ \left\{ {\rm conformations} \right\} } \exp \left[ -
\sum _{\alpha = 1}^n {\cal H}
\left( {\rm sequence, \ conformation}_{\alpha} \right) /T
\right] \ ,
\label{eq:symbol}
\end{eqnarray} where we explicitly mention the dependence of
the Hamiltonian (\ref{eq:GenHam}) on both sequence, which is
the same for all replicas $\alpha \in 1 \ldots n$, and
conformation, which is potentially different for different
replicas. Probability distribution over the set of sequences,
$ {\cal P}_{\rm sequence}$ is defined by the preparation
process and thus in our case can be written as
\begin{eqnarray} {\cal P}_{\rm sequence}  & \sim & 
\left[ p_{s_1}\cdot p_{s_2} \cdot \ldots \cdot p_{s_N} \right]
\times  
\nonumber \\  &
\times & \sum_{\rm target \ conformation} 
\exp \left[ {\cal H}^{p} \left( {\rm sequence, \ target \
conformation} \right) /T_p \right] \ ,
\label{eq:probab}
\end{eqnarray}  where we drop the normalization factor. In the
equation (\ref{eq:probab}), $p_s$ is the probability of
appearance of the monomer species $s$ (which is normally
controlled by the chemical potentials of components in the
monomer soup surrounding the preparation bath),
${\cal H}^{p}$ is Hamiltonian of the form (\ref{eq:GenHam})
except with the ``preparation'' matrix ${\Mat B}^{p}$ instead
of ${\Mat B}$ which controls folding through equation
(\ref{eq:symbol}). Accordingly, $T_p$ is the temperature at
which preparation process is performed.  

We stress that our approach is not restricted to any
particular target conformation. By contrast, we do average
over all possible (compact) target conformations (see equation
(\ref{eq:probab})), and thus our scheme picks up not just the
good sequences, but the pairs ``target conformation - 
sequence which is good for this target conformation,'' where
both terms are well adjusted to each other (see also the
discussion in Ref. \onlinecite{beer}).  This is a good match
for Imprinting, since we assume that some external field
chooses sequence-conformation pairs based upon matching with
the field
\cite{ExternalField}.  Indeed, this may be analogous to
protein  evolution, in  which nature chooses
sequence-conformation pairs not for any specific nature of the
conformation or sequence but for its functionality; this can
be viewed in physical terms as some external field affecting
the selection of sequence and conformation
\cite{ExternalField}.

\section{Free Energy of the Model} 

Inspection of the equations (\ref{eq:symbol}, \ref{eq:probab})
indicates that we can formally express  the weight
corresponding to the design process as an additional replica
labeled $0$ \cite{q2Design,Sharad,Design,Amit}: 
\begin{equation}
\left< Z^{n} \right>  = \sum_{\rm sequence} \prod_{I=1}^N
p_{s_I}  
\sum_{ \left\{ {\rm conformations} \right\} }  \exp \left[
\sum_{\alpha = 0}^n
\sum_{I \neq J =1}^N B^\alpha_{s_I,s_J}  \delta \left( {\Vec
r}_I^\alpha -  {\Vec r}_J^\alpha \right) / T_\alpha \right] \ ,
\label{eq:statsum}
\end{equation} where $B^{\alpha =0}_{ij} \equiv {\Mat B}^{p}$
is the matrix which expresses the interactions used for the
chain preparation (i.e. replica $\alpha=0$) and
$B^{\alpha>0}_{ij}
\equiv B_{ij}$ is the interaction matrix which governs folding
or renaturation.  Hereafter, conformations are given in terms
of position vectors ${\Vec r}_I^\alpha$ for each monomer
number $I$ and each replica $\alpha$. By the sum over
conformations we mean the sum in which the condition of chain
connectivity is strictly obeyed (technically this can be done
either in continuous form as Edwards \cite{Edwards} or in
discrete form like Lifshits \cite{Lifshits}).  

To facilitate averaging over the sequences, we define the
densities
\begin{equation}
\rho _i^\alpha ({\Vec R})=  \sum_I^N  \delta (s_I,i)
\delta({\Vec r}_I^\alpha - {\Vec R}) \ ,
\end{equation} then rewrite the exponent in equation
(\ref{eq:statsum}) as
\begin{equation}
\sum_{I \neq J =1}^N {B^\alpha_{s_I,s_J} \over T_\alpha}
\delta \left( {\Vec r}_I^\alpha -  {\Vec r}_J^\alpha \right) =
\int d{\Vec R}_1 d{\Vec R}_2 \sum_{i,j}^q \rho _i^\alpha
({\Vec R}_1) {B^\alpha_{i,j} \over T_\alpha} \delta({\Vec R}_1
-{\Vec R}_2) \rho _j^\alpha ({\Vec R}_2)
\end{equation}  and perform a Hubbard-Stratonovich
transformation on the quantity $\rho _i^\alpha ({\Vec R})$,
thus introducing the conjugate field
$\phi_i^\alpha({\Vec R})$.  We average over the sequence and
truncate the resulting exponent to ${\cal O}(\phi^2)$, which
yields (see the details in Ref. \onlinecite{Design}): 
\begin{eqnarray}
&& \left< Z^{n} \right> = \sum_{\rm conformations}  \int \! 
{\cal D} \left\{
\phi ({\Vec R}) \right\} \  \exp \left\{ \int d{\Vec R}  
\sum_{\alpha = 0}^n
\sum_{i}  \left[ p_i \rho ^{\alpha} ({\Vec R})
 \phi_i^\alpha({\Vec R}) \right]
\right. + \nonumber \\ 
&& \left.  \int d{\Vec R}_1 d{\Vec R}_2 
\sum_{\alpha,\beta = 0}^n \sum_{ij}  \left[  {1 \over 4} 
\left( {B_{ij}^\alpha \over T_\alpha } \right) ^{-1}
\delta^{\alpha \beta} \delta({\Vec R}_1,{\Vec R}_2)   +  
{1 \over 2}
\Delta_{ij} Q^{\alpha \beta}({\Vec R}_1,{\Vec R}_2) \right]
\phi_i^\alpha({\Vec R}_1) \phi_j^\beta({\Vec R}_2) \right\} 
\ ,
\end{eqnarray}
where we define the overall density $\rho_\alpha({\Vec R}) =
\sum_I^N \delta({\Vec r_I^\alpha} - {\Vec R}) = \sum_{i=1}^q
\rho _i^\alpha ({\Vec R}) $ and  the  replica overlap order
parameter
$Q_{\alpha \beta}({\Vec R}_1,{\Vec R}_2) =\sum_I^N
\delta({\Vec r_I^\alpha} - {\Vec R}_1)\delta({\Vec r_I^\beta} -
{\Vec R}_2)$.  Since the density is a single replica quantity
and we assume the chain as a whole is compressed, that is,
density is constant throughout  the globule, we simply take
$\rho_\alpha({\Vec R}) \equiv
\rho$.  Furthermore, using a variational argument,  it was
shown \cite{Sfatos,q2Design} that freezing occurs down  to
microscopic length scales, thus allowing to take 
$Q_{\alpha \beta}({\Vec R}_1,{\Vec R}_2) = \rho q_{\alpha
\beta} 
\delta({\Vec R}_1 - {\Vec R}_2)$, where the form of the
conformation correlator $q_{\alpha \beta}$ is found to be that
of a  Parisi matrix with one step symmetry breaking, with
either complete overlap ($q^{\alpha \beta}=1$) or no overlap
($q^{\alpha \beta}=0$). (This directly corresponds with the 
Random Energy Model
\cite{Derrida} introduced directly in previous heteropolymer
models
\cite{Wolynes}.)  This facilitates Gaussian integration over
the $\phi$  fields.  To write the result in even simpler form,
we can also include a conformation-independent constant by the
transformation
$\phi_i^{\alpha} \rightarrow 2 \sum_j
\left( \rho {\Mat B}^{\alpha}/T_\alpha \right)^{1/2}_{ij}  
\phi ^{\alpha}_j $     to get 
\begin{eqnarray}
\left< Z^{n} \right> &=& \sum_{\rm conformations}  
\left[ \int \! d \left\{ \phi \right\} \   \exp 
\left\{  \sum_{\alpha =
0}^n \sum_{ij}  \left[ 
\left( 2 { \rho \Mat B}^{\alpha}/T_\alpha \right)^{1/2}_{ij}
p_i   \phi_j^{\alpha} \right] + \right. \right. \nonumber \\   
&&  \left.
\left. \sum_{\alpha,\beta = 0}^n \sum_{ij}  \left[ 
\delta_{ij} \delta^{\alpha \beta}    + 
 2 \left( \left( {\Mat B}^{\alpha}/T_\alpha \right) ^{1/2} 
{\Mat \Delta} \left( {\Mat
B}^{\beta}/T_\beta \right)^{1/2} \right)_{ij} \rho 
q^{\alpha \beta} \right]
\phi_i^\alpha \phi_j^\beta  \  \right\} \ \right]^N \ .
\end{eqnarray}
where we use a hat to indicate that the object is matrix in
species space (i.e.  ${\Mat A } = A_{ij}$).  We evaluate this 
Gaussian integral, yielding the free energy
\begin{equation}
\left< Z^{n} \right> = \sum_{ \left\{ {\rm conformations}
\right\} } \exp\left[ E (q) \right] \ ,
\end{equation}  where the effective energy of the $n$ replica
system is given by
\begin{eqnarray}
{ E(q) \over N} &=&  {1 \over 2} \ln \det \left[ {\Mat I}
\delta^{\alpha \beta} +  
2 \rho \left( {\Mat B}^\alpha/T_\alpha \right)^{1/2} q^{\alpha \beta} {\Mat \Delta}
\left( {\Mat B}^\beta/T_\beta \right)^{1/2}  \right] +
\nonumber
\\  &&    { 1 \over \rho} \sum_{\alpha \beta} \left< {\vec
\rho} \left| \left( {\Mat B}^\alpha/T_\alpha \right)^{1/2}
 \left[ {\Mat I} \delta^{\alpha \beta} + 2 \rho 
\left( {\Mat B}^\alpha/T_\alpha \right)^{1/2} q^{\alpha \beta} {\Mat \Delta}
\left( {\Mat B}^\beta/T_\beta \right)^{1/2}
\right]^{-1} 
\left( {\Mat B}^\beta/T_\beta \right)^{1/2}
 \right| {\vec \rho } \right>  \ ,
\label{eq:energdoupr}
\end{eqnarray}
$\left<  \right\vert \cdots \left\vert  \right>$ denotes the
scalar product over species space, the determinant in the
first term is over species and replica space,  and the vector
${\vec \rho}$ is given by ${\vec \rho}_{i}^{\alpha} = p_i \rho
$. Note that the only remaining dependence on conformations
come through conformational correlators $q^{\alpha \beta}$. 
Given the particular structure of $q^{\alpha
\beta}$, effective energy (\ref{eq:energdoupr}) can be
expressed directly in terms of the number of replicas which
overlap with the target group $y$ and the size of a group $x$
for the remaining $n-y$ replicas divided into
$(n-y)/x$ groups.  Thus, we can simplify the expression for
effective energy (\ref{eq:energdoupr}) by removing  replica
dimensionalities, as is performed in Appendix A. This  also
allows one to write the entropy of the macrostate with given
$x$ and
$y$, as it is associated simply with grouping of replicas  $S
=N s [ y + (n-y) (x-1)/x] $. 
identical conformation down to microscopic scale related to
the volume $v$, there is an entropy loss of $s
\equiv
\ln(a^3/v)$ per monomer, where
$a$ is the distance between monomers and $v$ is the excluded
volume  
\cite{IndepenInt,Sfatos}.  This allows conversion from the sum
over conformations to a functional integral over   $Q^{\alpha
\beta}({\Vec R}_1,{\Vec R}_2)$, and even further, to
conventional integral over $x$ and $y$, which, in the mean
field approximation, can be further simplified to optimization
of the effective $n$-replica free energy 
\begin{eqnarray}
{ F(x,y) \over N} & = &  
{n-y \over 2 x} \left\{ \ln \det \left[ {\Mat I}  + 
2 x {\Mat \Delta} 
{\Mat B}/T  \right]  +
\left< {\vec p} \left| 2 x {\Mat B}^1/T \left[ {\Mat I}  + 
2 x {\Mat \Delta} 
{\Mat B}/T \right]^{-1} \right| {\vec p} \right> \right\}  
\nonumber \\
& + &{1 \over 2} \ln \det \left[ {\Mat I}  + 2 {\Mat \Delta} 
{\Mat B}^p/T_p + 2 y 
{\Mat \Delta} {\Mat B}/T  \right]  
\nonumber \\ 
& + &
\left< {\vec p} \left| \left[ {\Mat B}^p/T_p +  y {\Mat B}/T 
 \right]
\left[ {\Mat I}  + 2 {\Mat \Delta} {\Mat B}^p/T_p + 
2 y {\Mat \Delta} {\Mat 
B}/T \right]^{-1} \right| {\vec p} \right>  
\nonumber \\ &-&  s [ y + (n-y) (x-1)/x]
\label{eq:svobenerg}
\end{eqnarray}

\section{Analysis of the Free Energy and Phase Diagram}

The expression (\ref{eq:svobenerg}) is rather similar to what
we had in Ref.
\onlinecite{Design} while considering the model with identical
interactions for design and folding and, of course, it is
exactly reduced to the corresponding equation of that work
\cite{Design} when ${\Mat B} = {\Mat B}^{p}$. Furthermore,
this expression implies the same structure of phase diagram,
with the same three globular phases: random, frozen, and
target. (We remind the reader, that overall collapse of the
chain is the necessary pre-condition of our approach, and thus
globule-to-coil phase transition falls outside of the
framework of the present study). To see the structure of phase
diagram, we first look at the allowed variations of the order
parameters $x$ and $y$. 

For simplicity, we consider here only small $s$ regime. In
this case, freezing transitions, which are the main topic of
our interest here, occur when $B$ is (in a reasonable sense)
also small. Indeed, freezing phase transitions result
physically from the competition between energetic and entropic
parts of free energy (\ref{eq:svobenerg}), where energetic
part favors gathering of replicas into groups while entropic
part favors diversity of replicas. For energy to be
competitive to an entropy when $s$ is small, $B$ must be small
as well. This allows one to simplify equation
(\ref{eq:svobenerg}) truncating it to quadratic order in $B$.

As $y$ is the number of replicas whose conformation coincides
with the target conformation, this value must be in between of
$0$ and $n$. What is relevant in replica approach is $n
\rightarrow 0$ limit, and, moreover, only the terms which are
linear in $n$ are to be considered (because higher order terms
disappear in the main equation
$\left< \ln Z \right> = \lim_{n \rightarrow 0} \left( \left< Z
^n \right > - 1
\right) /n$). Accordingly, since $0
\leq y \leq n$, we must  linearize the free energy in $y$ as
well 
\cite{q2Design,Design}. This leads to further simplification of
(\ref{eq:svobenerg}):
\begin{eqnarray}
F & = &  
{\rm Tr} \left[ (n-y)  \left\{ 
{\Mat \Delta} {\Mat B}/T - x {\Mat \Delta} {\Mat B} 
{\Mat \Delta} {\Mat B}/T^2
+ {\Mat P} {\Mat B}/T - 2 x {\Mat P} {\Mat B} 
{\Mat \Delta} {\Mat B}/T^2 
\right\} \right.
\nonumber \\ 
&& + 
{\Mat \Delta} {\Mat B}^p/T_p +  y  {\Mat \Delta} {\Mat B}/T 
- 2 y {\Mat \Delta} {\Mat B}^p {\Mat \Delta} {\Mat B}/T T_p
-   {\Mat \Delta} {\Mat B}^p {\Mat \Delta} {\Mat B}^p/T_p^2
\nonumber \\ 
&& + \left.
{\Mat P} {\Mat B}^p/T^p + y {\Mat P} {\Mat B}/T -
2 \left( {\Mat P} {\Mat B}^p   {\Mat \Delta} 
{\Mat B}^p/T_p^2 + 
       y {\Mat P} {\Mat B} {\Mat \Delta} 
{\Mat B}^p/T T_p +   
       y {\Mat P} {\Mat B}^p  {\Mat \Delta} 
{\Mat B}/T T_p \right) 
\right]  
\nonumber \\ 
&&
- T N s [ y + (n-y) (x-1)/x]
\label{eq:iiapprox}
\end{eqnarray}
where $P_{ij} \equiv p_i p_j$.

While $y$ describes breaking of the symmetry between $n$
replicas due to their attraction to the target replica labeled
$0$, $x$ describes spontaneous symmetry breaking.  When we
have integer number of replicas, $n$, clearly, $1 \leq x \leq
n$:
$x$ cannot be smaller than unity, because it is the number of
replicas in the group. When $n \rightarrow 0$, the logic about
the number of replicas in the group is not applicable any
more, but it is natural to think that formal inequalities for
$x$ just simply flip signs: $n \leq x \leq 1$. With this in
mind, we optimize free energy (\ref{eq:iiapprox}) with respect
to $x$ yielding the equation which determines $x$:
\begin{eqnarray}
s = { x^2 \over T^2} {\rm Tr} \left[ 
 {\Mat \Delta} {\Mat B} {\Mat \Delta} {\Mat B}
 + 2  {\Mat P} {\Mat B} {\Mat \Delta} {\Mat B} \right]
\label{eq:sRand}
\end{eqnarray}
Note, that this equation does not involve either $T_p$ or
${\Mat B}^{p}$ and thus it does not depend on preparation
process. This has clear physical meaning. Namely, this
reflects the behavior similar to that of REM, because the
designed sequence behaves precisely as a random one in  all
the conformations except for the target conformation.

At this point, it is useful to introduce the following matrix 
``cumulants'': 
\begin{eqnarray}
\left< A \right>_c & \equiv & \sum_{i,j} p_i p_j A_{ij} =
 {\rm Tr} \left( {\Mat P} {\Mat A} \right)
\nonumber \\ 
\left< A B \right>_c & \equiv & \sum_{i,j} p_i p_j A_{ij}  
B_{ij}  - \left< A \right>_c \left< B \right>_c 
   = {\rm Tr} \left( {\Mat \Delta} {\Mat A} {\Mat \Delta}
{\Mat B}
          + 2  {\Mat P} {\Mat A} {\Mat \Delta} {\Mat B} \right) 
\label{eq:cumulantsdef}
\end{eqnarray}
where ${\Mat A}$ and ${\Mat B}$ are arbitrary matrices.

From the above, we can easily find the equation for the
freezing temperature for random  sequences. Indeed, freezing
occurs when replicas start to group, thus spontaneously
breaking  the permutation symmetry. This happens when $x=1$.
Therefore,  freezing temperature is given via the relation
\begin{eqnarray}
T_f^2 =  \left< B B \right>_c /s
\label{eq:Tf}
\end{eqnarray}
In other words, the freezing temperature is given by the
variance of the  renaturation interaction matrix
\cite{RandHet}.  Note that this is a transition to a unique
ground state which is not necessarily (and most likely not)
the target conformation: we call this phase the {\em frozen
phase} and we call the high temperature disordered phase in
which there is no form of freezing, i.e. many conformations
dominate equilibrium, the {\em random} phase.

To examine freezing to the target conformation, we must examine
the conditions at which $y>0$.  Since $y$ varies from $0$ to
$n$, what  has physical meaning in the $n \rightarrow 0$ limit
is only the linear  in $y$ term of free energy. Therefore,
free energy optimum corresponds to  either $y=0$ (non-target
phase), or to $y=n$ (target phase). To find the  corresponding
critical temperature, we must examine the slope of the free
energy at the point $y=0$ to determine whether $y=0$ or $y=n$
is the stable solution \cite{Design}. The condition
``slope''$=0$ yields the  relationship:
\begin{eqnarray}
s & = &   {\rm Tr}  \left[
  x \left( { {\Mat \Delta} {\Mat B}  \over T} 
           { {\Mat \Delta} {\Mat B}^p \over T_p} + 
      2 { {\Mat P} {\Mat B} \over T} { {\Mat \Delta} 
{\Mat B}^p \over T_p} +   
{ {\Mat \Delta} {\Mat B}^p \over T_p} { {\Mat \Delta} 
{\Mat B}  \over T} + 
2 { {\Mat P} {\Mat B}^p \over T_p} { {\Mat \Delta} 
{\Mat B}  \over T} \right)
- {x^2 \over T^2} \left( {\Mat \Delta} {\Mat B}
 {\Mat \Delta} {\Mat B}
 + 2  {\Mat P} {\Mat B} {\Mat \Delta} {\Mat B} \right) 
\right]  
\nonumber \\
& = & {2 x  \over T_p T}\left<  B^p B  \right>_c - {x^2 \over T^2} \left< B B  
\right>_c
\label{eq:sDes}
\end{eqnarray}
This equation defines the phase boundary of the {\em target
phase}, in which the system freezes to the target conformation.

We combine eqns (\ref{eq:sRand}-\ref{eq:sDes}) to get the
boundary of the target phase (i.e. the preparation temperature
$T_p$ which separates the target phase from the random and
frozen globule phases). To write the result, it is convenient
to define  formally the value of $T_{pf}$ according to the
equation
\begin{eqnarray}
T_{pf}^2 =  \left< B^p B^p \right>_c /s
\label{eq:Tpf}
\end{eqnarray}
similar to eq (\ref{eq:Tf}) except it includes preparation
matrix $\Mat B ^p$ instead of $\Mat B$. Physically, $T_{pf}$
is the temperature point at which random heteropolymer would
undergo freezing transition provided its conformations are
governed by $\Mat B ^p$ interaction matrix; $T_{pf}$ gives a
natural scale for $T_p$:
\begin{equation}
{T_p^{(cr)} \over T_{pf} } = \cases{ 
g {2  T/T_f \over (1 + T^2/T_f^2)} & for $T \geq T_f$ \cr 
g 						          & for $T \leq T_f$
} \ ;
\label{eq:Ttar}
\end{equation}
as far as the small $s$ limit is concerned, this can be also
rewritten in terms of $T_{\rm tar}$, acting temperature at
which random-to-target phase transition occurs:
\begin{equation}
{T_{\rm tar} \over T_f} = 1 + \left[ 1- {T_p  \over g T_{pf} } \right] ^{1/2} \ .
\label{eq:Ttarold}
\end{equation}
This is the previously obtained result for the transition to
the target phase \cite{Design}, except with the inclusion of
a  factor $g$, which is defined as
\begin{equation}
g \equiv \left< B^p B \right>_c/ \sqrt{ \left< B B \right>_c \left< B^p B^p \right>_c } \ .
\label{eq:definitiong}
\end{equation}
To understand the meaning of $g$, it must be noted first of
all that $g$ can be treated as scalar product, $g = \cos
\theta$, and thus $-1 \leq g \leq 1$.  This factor gives the
degree of correlation between the elements of the two
matrices, $\Mat B$ and $\Mat B ^p$.  If the two matrices are
the same (i.e. completely correlated)
$g=1$.  The region  $0 < g < 1$ corresponds to somewhat
lesser, but still positive, degree of correlation; $g=0$ means
that matrices are statistically independent; $-1 < g < 0$
corresponds to some anti-correlation; finally, $g=-1$ means
absolute anticorrelation (each pair of monomers which is
supposed to be attractive in $\Mat B$, is repulsive in $\Mat B
^p$, and vice versa, etc). To see this, it is helpful to note,
that the definition of matrix cumulants has the property that
$\left< B^1 B^2 \right>$, where $\Mat B ^1$ and $\Mat B ^2$
are both either $\Mat B$ or $\Mat B ^p$, does not change upon
adding a constant to all matrix elements (this can be easily
proven given that $\sum _i p _i =1$); this allows to define $
b_{ij} ^x   = B^x_{ij} - \left< B^x
\right>_c $ and rewrite eq (\ref{eq:definitiong}) as $g =
\left< b^p b \right>_c/ \sqrt{
\left< b b \right>_c \left< b^p b^p \right>_c }$.  It is now
seen, that, for example,
$g=-1$ corresponds to $b_{ij} = - b^p_{ij}$.   

According to eq (\ref{eq:Ttar}), the value $T_p^{(cr)}$  is
proportional to $g$.  At
$g=1$, we recover the result for $\Mat B^p = \Mat B$
\cite{Design}. Appearance of positive, but not absolute
correlation ($0 < g < 1$) has simple graphical meaning on the
phase diagram, Fig. 1 --- it leads to affine deformation of
the boundary of target phase region on the phase diagram. At
$g=0$, the target region disappears, and it does not exist at
$g <0$. This is  clear, because when matrices are
anticorrelated, ``design'' does not help, but rather destroys
the chances of polymer to fold into  desirable conformation. 
Thus, the correlation between matrices, given by the factor
$g$, is the measure of the proximity between interaction
matrices.

\section{Discussion}

By performing explicit calculations for the freezing
transition of heteropolymers with different matrices for
design and renaturation, we have found three phases: {\em
random}, in which many conformations dominate equilibrium;
{\em frozen}, where the polymer freezes to a single
conformation other than the target conformation; and {\em
target}, in which the polymer freezes to the target
conformation. In the flexible chain limit, for the case where
the design and renaturation matrices are different, the
effective critical selective temperature for renaturation to
the target phase becomes modified by a factor from the
normalized correlation between the matrices ($T_p^{(cr)}
\rightarrow T_p^{(cr)} g$).  For complete correlation, $g=1$. 
For differences in the design and renaturation matrix ($g<1$),
special measures must be undertaken in order to keep the
system in the target phase; otherwise, there is no possibility
to obtain renaturation to the correct target conformation.

To understand better the meaning of the result obtained,
consider that proteins have been indeed ``designed'' according
to one of the theoretical models
\cite{ShakDesign,LongImprint,Dill,Wolynes}.  This means, that
they were prepared under the interactions $\Mat B^p$ at some
temperature $T_p < T_{pf}$, and now they ``work'' at some
other temperature $T$, such that $ T_{pf} < T < T_{\rm tar}$.
It is worth stressing, that their ``work'' is governed by
their natural interactions, that is, by the same matrix $\Mat
B ^p$ as was supposedly used for ``design.''  We take now some
other artificial matrix $\Mat B$ and try (for example, by
means of computer simulation) to recover the correct
renaturation. In terms of our phase diagram, Fig. 1, correct
renaturation occurs when and only when the system remains in
the target phase.  This is illustrated graphically in  Fig. 1:
the ``Natural'' phase diagram is presented there with solid
lines, and real proteins are supposedly represented by the
point within target phase region.  Phase diagrams of a couple
of artificial systems, with sequences designed by natural
matrix
$\Mat B ^p$ and conformations governed by mistaken matrices 
$\Mat B$, are shown with dashed lines; a greater degree of
errors in the potentials push the phase boundary to the left,
determined by the value of the factor $g$ (defined by
(\ref{eq:definitiong}) ). In the example illustrated in the
figure, the representative point for matrices with $g>0.95$
remain within the target phase region, while those with
$g<0.95$ do not possess this property. In the first case,
correct renaturation can be recovered, in the second case this
is impossible.  We conclude, that correct renaturation is
possible when the degree of correlation is sufficient between
$\Mat B ^p$ and $\Mat B$, namely, when
\begin{equation}
g > g^{\ast} \ \ \ \ \ 
{\rm (the \ condition \ for \ correct \ renaturation)} \ ,
\end{equation}
where $g^{\ast}$ is defined from the condition that the
boundary $T_{\rm tar}$, given by the equation
(\ref{eq:Ttarold}), goes through the given point $(T_p , T)$:
\begin{equation}
g^{\ast} = {T_p \over T_{pf}} + 
\left[ {T \over T_f} - 1 \right]^2 \ .
\label{eq:gstar}
\end{equation}
This can be also instructively rewritten as
\begin{equation}
g^{\ast} = 1 + {T_p^{(cr)} - T_p \over T_{pf}}  \ .
\label{eq:gstar}
\end{equation}
Note, that the ratio $T_p / T_{pf}$ can serve as a measure of
degree of selection of sequences: smaller values of  $T_p /
T_{pf}$ correspond to stronger selection of sequences. At the
same time, $\left[T_p^{(cr)} - T_p \right] / T_{pf}$ measures
the degree of necessary selection at the given actual
temperature, $T$ (because $T_p^{(cr)}$ depends on $T$,
equation (\ref{eq:Ttar})).  We conclude, that minimal
``correctness'' of interaction matrix, $g^{\ast}$, is defined
by the degree of selection, or optimization, of the set of
real sequences: the better they have been optimized, the more
stable is their renaturation with respect to the mistakes in
interactions.

Speaking about the numbers involved in the problem, we have to
stress that the information available is by far insufficient
to make any solid statements.  To get some rough idea, we can
proceed in the following way.  When extracting a matrix of
species-species energies for proteins from the statistics of
protein data bank, such as the Miyazawa and Jernigan (MJ)
matrix \cite{Jernigan}, what one obtains is actually
$B_{ij}^{\rm MJ} = B_{ij}^p/T_p$ (see  Ref. \onlinecite{Gutin}
and Appendix B).  

Since from equation (\ref{eq:Tf}) we have, $\left< B_{ij}^2
\right>_c^{1/2} = s^{1/2} T_f$, then the variance of the MJ
matrix yields $\left< (B_{ij}^{\rm MJ})^2 \right>_c^{1/2} =
s^{1/2} T_{pf}/T_p$; therefore, with the knowledge of the MJ
matrix ($\left< (B_{ij}^{\rm MJ})^2 \right>_c \approx 2.0)$
and the flexibility of proteins
$s$ ($s \approx 1.6$), we  also arrive at $T_p/T_{pf} \approx
0.9$.  It is also independently hypothesized \cite{Goldstein}
that the ratio of the ``folding'' to the ``glass'' temperature
should be about $1.6$.  Without going into the arguments of
the work Ref. \onlinecite{Goldstein}, we can, quite
arbitrarily, identify ``folding'' temperature with $T_{\rm
tar}$ and ``glass'' temperature with $T_{pf}$; by doing so, we
obtain $T_{\rm tar}/T_f \approx 1.6$, which, in view of the
equation (\ref{eq:Ttar}), yields a similar estimate for the
degree of optimization
$T_p/T_{pf} \approx 0.9$.  We conclude, that a conservative
estimate of $g^{\ast}$ is likely to be about  $g^{\ast}
\approx 0.95$: correct recovery of the native state require $g
> 0.95$, if $g<0.95$ chances of correct renaturation are slim. 

Yet another, though purely algebraical, aspect of the problem
is which types and values of errors in determination of
interactions lead to a particular value of $g$ factor, such
as, for instance, $0.95$ mentioned above.  We first of all
note that neither additive ($B_{ij} = B^p_{ij} + B_0$) nor
multiplicative ($B_{ij} = \beta B^p_{ij}$) systematic errors
do not contribute at all, $g=1$ in both cases (as well as in
the ``combined'' case $B_{ij} = \beta B^p_{ij} + B_0$). This
is clear physically because these kinds of errors contribute
to homopolymer terms  only and do not affect selectivity of
interactions of monomers to one another. On a more formal
level, additive constants does not change second moments of
matrices defined according to eq~(\ref{eq:cumulantsdef}), and
multiplicative constant obviously does not affect the value of
$g$~(\ref{eq:definitiong}). To get an idea about random
mistakes, we examine the case where the renaturation matrix is
the design matrix with some normally distributed noise 
$\eta_{ij}$: 
$B_{ij}  = B^p_{ij}( 1 + \eta_{ij} ) $,  where  ${\cal P}
(\eta_{ij}) \propto \exp [ - \eta_{ij}^2/ \sigma^2 ]$.  We can
average the $g$ factor over the noise to get
\begin{equation}
\bar{g} = 
{ \left< B_{ij}^p \left[ B_{ij}^p (1 + \eta_{ij}) \right] 
\right> \over 
\left[ \left< \left[ B_{ij}^p ( 1 + \eta_{ij}) \right]^2
 \right> \right]^{1/2}
} =\left(1  + \sigma^2 \right)^{-1/2} 
\label{eq:gav}
\end{equation}
This gives $\sigma \approx 10\%$.   More complicated random,
systematic and mixed errors may be interesting to model and
this can be easily accomplished within this formalism but the
results are dependent of the specific nature of these errors
and are therefore not within the more general scope of this
paper.

It is worth making very clear that this error limit is
independent of the length of the polymer.  Previous
calculations \cite{Bryngelson} have made estimates which are
directly based upon $N$ (ie. the error must be small compared
with
$1/\sqrt{N}$). This reflects fundamental difference of our
approach from that of the Ref.
\onlinecite{Bryngelson};  even more, this reflects the
difference between questions studied.  In the work Ref.
\onlinecite{Bryngelson}, calculations were performed for
random heteropolymers, neither any kind of design nor the
principle of minimal frustration was imposed; accordingly, the
question studied was in fact about the possibility to
reconstruct the randomly chosen conformation of frozen globule
phase. In the ensemble of random sequences, the ones with very
stable ground state are very (exponentially) rare, thus, it is
not surprising that these ground states are typically very
unstable with respect to error-based renaturation, especially
for long chains. By contrast, our treatment is a comparison
between the types of freezing (to the target or some random
conformation).  This is therefore independent of the length of
the polymer chain and essentially of a different nature than
that of Ref.
\onlinecite{Bryngelson}.  Furthermore, within our formalism,
the transition in
$y$, the number of replicas in the target group, is first
order; therefore,  in the framework of our approach one cannot
discuss the ``degree'' of renaturation in terms of a given
percentage of correct contacts: in thermodynamic equilibrium,
and in very long chain (thermodynamic limit) either there is
renaturation to the target conformation or folding to  some
entirely different conformation.

Within the framework of our formalism, the Independent 
Interaction Model \cite{IndepenInt} can be recovered by
addressing the limit $q
\rightarrow N$ and assuming that $B$ is a normally distributed
matrix; in this case eq (\ref{eq:Tf}) agrees with the results
of more direct calculations of this model \cite{IndepenInt}. 
The error limits in this approximation are derived in exactly
the same manner as (\ref{eq:gav}).  This is not surprising as,
in fact, the validity of the approximation of taking the free
energy to ${\cal O}(B^2)$ in eq (\ref{eq:iiapprox}) is similar
to that of the Independent Interaction Model \cite{Design}: we
assume that the effective flexibility $s$ of the polymer is
small.  However, our treatment allows corrections to this
approximation to be systematically derived.

In conclusion, starting from the most general Hamiltonian
involving short range binary heteropolymeric interactions, we
have derived what measure is used to compare differences in
interaction potentials and the limits in which renaturability
to the target conformation is still allowed.  Simple estimates
of normally distributed error indicates that even conservative
estimates leave room for $10\%$ error in potentials.  Using
our formalism, one can make a more informed estimate based
upon more precise knowledge of the form of errors involved,
i.e. the correlations of errors in the matrix.

\bigskip

\centerline{\bf ACKNOWLEDGEMENTS}

The work was supported by NSF (DMR 90-22933)  and NEDO of
Japan. VSP acknowledges the support of an NSF Fellowship. AYG
acknowledges the support of Kao Fellowship.

\newpage
\appendix

\section{Simplification of Equation (7)}

We will a slightly different notation from the rest of the
paper to facilitate calculations: we eliminate indices and
simply give the dimensionality of the operators explicitly,
eg. we label $\Delta_{ij}$ as 
${\Mat \Delta}^{(q)}$ since it is a $q \times q$ dimensional
matrix.

We perform the simplification of the elimination of replicas
though several steps:
\begin{enumerate}
\item{${\Mat q}$ is of well-known one-step replica symmetry
breaking shape, with one distinct group of $y+1$ replicas and
$(n-y)/x$ groups of $x$ replicas each.}
\item{${\Mat M} =  {\Mat I}  + 2 \rho {\Mat q}  \otimes  {\Mat
\Delta} \  {\Mat B}$ can be viewed as $(n+1) \times (n+1)$
block matrix in replica space, with each matrix element being
$q \times q$ matrix in species space. This block matrix is of
the same structure as ${\Mat q}$, with one $(y+1) \times
(y+1)$ super-block and  $(n-y)/x$ of $x \times x$
super-blocks.}
\item{The determinant in the first term in free energy is
decomposed into the product of determinants of super-blocks. }
\item{Vector ${\vec \rho}$ is composed of $n+1$ ``blocks''
$p_i$, thus making the second term in free energy the sum of
independent contributions from the groups of replicas. Along
with previous, this means that different groups of replicas do
hot interact and this is why they contribute independently to
the free energy.}
\item{Effective replica energy $E$ is now presented in the form
\begin{equation}
{ E(x,y) \over N} = \epsilon _y + {n-y \over x } \epsilon_x \ ,
\end{equation} 
where $\epsilon _y$ and $\epsilon _x$ are the (independent) contributions from the corresponding groups of replicas. (Note, that replica entropy is also of the same form).}
\item{Both $\epsilon _y$ and $\epsilon _x$ have almost the same form as $E$ (\ref{eq:energdoupr}), except simpler matrix ${\Mat {\tilde q}}$, with all matrix elements $1$, appears instead of ${\Mat q}$:
\begin{eqnarray}
\epsilon_z  & = &  {1 \over 2} \ln \det \left[ {\Mat I}^{(zq)}  +  2 \rho {\Mat {\tilde q}}^{(z)}  \otimes  {\Mat \Delta}^{(q)} \  {\Mat B}^{(zq)} \right] + \nonumber \\ & + &
{1 \over \rho} \left< {\vec \rho} \left|  {\Mat B}^{(zq)} \left[ {\Mat I}^{(zq)}  + 2 \rho {\Mat {\tilde q}}^{(z)}  \otimes  {\Mat \Delta}^{(q)} \  {\Mat B}^{(zq)} \right]^{-1} \right| {\vec \rho} \right>  \ ,
\label{eq:onegroupenergy}
\end{eqnarray}
where $z$ is either $x$ or $y+1$, i.e., the number of replicas in the group. }
\item{To simplify first term (with determinant), we define rotation unitary operator 
\begin{equation}
{\Mat {\cal R}}^{(z)}_{\alpha \beta} = {1 \over \sqrt{z} } \exp \left[ {2 \pi i \over z} (\alpha -1) (\beta -1) \right] \ \ \ \ \ \ \ \ \ \ \ \  1 \leq \alpha, \beta \leq z \ .
\end{equation}
It is easy to check that this operator transforms  ${\Mat {\tilde q}}^{(z)}$ into diagonal form, where one diagonal matrix element is $1$, while all others are $0$:
\begin{equation}
{\Mat {\cal R}}^{(z)}{\Mat {\tilde q}}^{(z)}\left({\Mat {\cal R}}^{(z)}\right)^{-1} = {\Mat \lambda}^{(z)} \ , {\rm \ \ \ \  where \ \ \ \ \ \ } \ \ \ \ \ {\Mat \lambda}_{\alpha \beta} = z \delta_{\alpha 1}   \delta_{1 \beta} \ .
\label{eq:lambda}
\end{equation}
We define also ${\Mat {\cal R}}^{(zq)}={\Mat I}^{(q)} \otimes {\Mat {\cal
R}}^{(z)}$ and note that the determinant is not changed upon rotation. We write
\begin{eqnarray}
&& \det  \left[ {\Mat I}^{(zq)}  +  2 \rho {\Mat {\tilde q}}^{(z)}  \otimes  {\Mat \Delta}^{(q)} \  {\Mat B}^{(zq)} \right]  =  \nonumber \\ 
& = & \det \left[ {\Mat {\cal R}}^{(zq)} \right]  \det \left[ {\Mat I}^{(zq)}  +  
						2 \rho {\Mat {\tilde q}}^{(z)}  \otimes  {\Mat \Delta}^{(q)} \  {\Mat B}^{(zq)} \right]  \det \left[ {\Mat {\cal R}}^{(zq)} \right]^{-1}  \nonumber
\\   
& = & \det \left[ {\Mat I}^{(zq)}  +  2 \rho \left( {\Mat {\cal R}}^{(zq)} \right) {\Mat {\tilde q}}^{(z)}  \otimes  {\Mat \Delta}^{(q)} \  {\Mat
B}^{(zq)} \left( {\Mat {\cal R}}^{(zq)} \right)^{-1}\right]  \nonumber \\  
& = & \det \left[ {\Mat I}^{(zq)}  +  2 \rho \left( {\Mat {\cal R}}^{(z)}
\right) {\Mat {\tilde q}}^{(z)} \left( {\Mat {\cal R}}^{(z)} \right)^{-1} \otimes  {\Mat \Delta}^{(q)} \ \left( {\Mat {\cal R}}^{(zq)} \right) {\Mat
B}^{(zq)} \left( {\Mat {\cal R}}^{(zq)} \right)^{-1}\right]  \nonumber \\ 
& = & \det \left[ {\Mat I}^{(zq)}  +  2 \rho  {\Mat \lambda}^{(z)} \otimes 
{\Mat \Delta}^{(q)} \ \left( {\Mat {\cal R}}^{(zq)} \right) {\Mat B}^{(zq)} \left( {\Mat {\cal R}}^{(zq)} \right)^{-1}\right] \ . 
\end{eqnarray} 
As  ${\Mat B}^{(zq)}$ is diagonal in replica space, ${\Mat B}^{(zq)} = {\Mat B}^{(q)}_{\alpha} \delta _{\alpha \beta}$, we have 
\begin{eqnarray}
&& \left( \left( {\Mat {\cal R}}^{(zq)} \right) {\Mat B}^{(zq)} \left( {\Mat {\cal R}}^{(zq)} \right)^{-1} \right)_{\alpha \beta} = \sum_{\gamma \delta} {\Mat {\cal R}}_{\alpha \gamma} {\Mat B}^{(q)}_{\gamma} \delta _{\gamma \delta} \left( {\Mat {\cal R}}^{(zq)} \right)^{-1}_{\delta \beta} = \nonumber \\
& = & {1 \over z} \sum_{\gamma} \exp \left[ { 2 \pi i \over z} (\alpha - \beta) (\gamma - 1) \right]  {\Mat B}^{(q)}_{\gamma}  \ .
\end{eqnarray} 
Taking into account the simple structure of ${\Mat \lambda}$
(\ref{eq:lambda}), we arrive at
\begin{equation}
{\Mat \lambda}^{(z)} \otimes  {\Mat \Delta}^{(q)} \ \left( {\Mat {\cal R}}^{(zq)} \right) {\Mat B}^{(zq)} \left( {\Mat {\cal R}}^{(zq)} \right)^{-1} = \delta_{\alpha 1}  \sum_{\gamma} \exp \left[ { 2 \pi i \over z} (1 - \beta) (\gamma - 1) \right] {\Mat \Delta}^{(q)} {\Mat B}^{(q)}_{\gamma}
\end{equation}
}
\item{First consider a non-target group of $z=x$ replicas. In this group, all the replicas are identical meaning that ${\Mat B}^{(q)}_{\gamma}={\Mat B}^{(q)}$ does not depend on replica index $\gamma$. This yields 
\begin{equation}
{\Mat \lambda}^{(z)} \otimes  {\Mat \Delta}^{(q)} \ \left( {\Mat {\cal R}}^{(zq)} \right) {\Mat B}^{(zq)} \left( {\Mat {\cal R}}^{(zq)} \right)^{-1} = x \delta_{\alpha 1} \delta_{1 \beta } {\Mat \Delta}^{(q)} {\Mat B}^{(q)}
\end{equation}
and thus 
\begin{equation}
 \det \left[ {\Mat I}^{(xq)}  +  2 \rho  {\Mat {\tilde q}}^{(x)}  \otimes  {\Mat \Delta}^{(q)} \  {\Mat B}^{(xq)} \right] = \det \left[  {\Mat I}^{(q)}  +  2 \rho  x {\Mat \Delta}^{(q)} \  {\Mat B}^{(q)} \right]
\label{eq:xdeter}
\end{equation}
}
\item{Consider now target group of $z=y+1$ replicas. In this case, ${\Mat B}^{(q)}_{\gamma}={\Mat B}_p^{(q)}$ for $\gamma = 1$ and ${\Mat B}^{(q)}_{\gamma}={\Mat B}^{(q)}$ otherwise. We write therefore
\begin{eqnarray}
{\Mat I}^{(zq)} &+& {\Mat \lambda}^{(z)} \otimes  {\Mat \Delta}^{(q)} \ \left( {\Mat {\cal R}}^{(zq)} \right) {\Mat B}^{(zq)} \left( {\Mat {\cal R}}^{(zq)} \right)^{-1} = \nonumber \\ &=& {\Mat I}^{(q)} \delta_{\alpha \beta} + \delta_{\alpha 1} {\Mat \Delta}^{(q)} \left( {\Mat B}_p^{(q)} - {\Mat B}^{(q)} \right) + (y+1) \delta_{\alpha 1} \delta_{1 \beta } {\Mat \Delta}^{(q)} {\Mat B}^{(q)} \ .
\end{eqnarray}
This is the block matrix of the peculiar form such that only upper block is non-zero in the first column; for that reason, its determinant is equal to the product of determinants of diagonal blocks (see Lemma 1). Thus,
\begin{equation}
 \det \left[ {\Mat I}^{\left( (y+1)q \right) }  +  2 \rho  {\Mat {\tilde q}}^{(y+1)}  \otimes  {\Mat \Delta}^{(q)} \  {\Mat B}^{\left( (y+1)q \right) }  \right] = \det \left[  {\Mat I}^{(q)}  +  2 \rho   {\Mat \Delta}^{(q)} \ \left( y {\Mat B}^{(q)} + {\Mat B}^{(q)}_p \right) \right] \ .
\label{eq:ydeter}
\end{equation}
}
\item{As to the second term in $\epsilon_z$ (\ref{eq:onegroupenergy}), it is easily computed using Lemma 2. Indeed,  ${\Mat B}^{\left( (y+1)q \right)}$ is block diagonal matrix with one block ${\Mat B}_p^{\left(q \right)}$ and $y$ others ${\Mat B}^{\left(q \right)}$. On the other hand,  $ {\Mat {\tilde q}}^{(y+1)}  \otimes  {\Mat \Delta}^{(q)}$ is the block matrix with every block being the same ${\Mat \Delta}^{(q)}$. Therefore, the matrix in question, $\left[ {\Mat I}^{(zq)}  + 2 \rho {\Mat {\tilde q}}^{(z)}  \otimes  {\Mat \Delta}^{(q)} \  {\Mat B}^{(zq)} \right]$, is exactly of the form ${\Mat V}_{{\Mat g},{\Mat h}}^{(z)}$ form, where ${\Mat g} = {\Mat \Delta}^{(q)}{\Mat B}^{(q)}_p$ and ${\Mat h} = {\Mat \Delta}^{(q)}{\Mat B}^{(q)}$. Using block matrix multiplication rule, it is easy to compute ${\Mat B}^{\left( (y+1)q \right)} {\Mat V}^{(y+1)}_{{\Mat e},{\Mat f}}$ (see Lemma 2) and then to use the result of Lemma 3. This finally gives 
\begin{eqnarray}
&& {1 \over \rho} \left< {\vec \rho} \left|  {\Mat B}^{(zq)} \left[ {\Mat I}^{(zq)}  + 2 \rho {\Mat {\tilde q}}^{(z)}  \otimes  {\Mat \Delta}^{(q)} \  {\Mat B}^{(zq)} \right]^{-1} \right| {\vec \rho} \right> = \nonumber \\ 
&& \ \ \ \ \ \ \ = \rho \left< {\vec p} \left| \left( {\Mat B}^{(q)}_p + y {\Mat B}^{(q)} \right) \left[ {\Mat I}^{(q)} + 2 \rho y {\Mat \Delta}^{(q)} {\Mat B}^{(q)} + 2 \rho {\Mat \Delta}^{(q)} {\Mat B}^{(q)}_p \right]^{-1} \right| {\vec p} \right>
\end{eqnarray}
}
\item{Similar expression for a non-target group of $x$ replicas can be derived from here by  
formally putting $ {\Mat B}^{(q)}_p \rightarrow  {\Mat B}^{(q)}$ and $y+1 \rightarrow x$, 
this gives 
\begin{equation}
 {1 \over \rho} \left< {\vec \rho} \left|  {\Mat B}^{(zq)} \left[ {\Mat I}^{(zq)}  + 2 \rho {\Mat {\tilde q}}^{(z)}  \otimes  {\Mat \Delta}^{(q)} \  {\Mat B}^{(zq)} \right]^{-1} \right| {\vec \rho} \right>  = \rho \left< {\vec p} \left| x {\Mat B}^{(q)}  \left[ {\Mat I}^{(q)} + 2 \rho x {\Mat \Delta}^{(q)} {\Mat B}^{(q)}  \right]^{-1} \right| {\vec p} \right>
\end{equation}
}
\end{enumerate}

\centerline{\bf Lemma 1.}

Consider an auxiliary problem of the matrix


This is block matrix, where ${\Mat g}$ is $q \times q$ matrix
and ${\Mat I}$ is identity matrix of the same size $q \times
q$. The question is to find the determinant of this matrix. 

It can be shown by expansion over the elements of the first
column, then over the elements of the first column of the
remaining minor, and by repeating this operation $q$ times,
that
\begin{equation}
\det \left[ {\Mat U}^{(z)}_{{\Mat g}} \right] = \det {\Mat g}
\end{equation} independently of the blocks placed in the
upper-right triangle (shown conventionally with question
marks).

\centerline{\bf Lemma 2.}

Consider another auxiliary problem of the following block
matrix:

\vspace{-.0cm}
\vspace{0cm}

Here ${\Mat g}$ and ${\Mat h}$ are matrices $q \times q$, they
generally do not commute to each other. ${\Mat I}$ is identity
matrix of the same size $q \times q$. Total size of the block
matrix ${\Mat V}_{{\Mat g},{\Mat h}}^{(z)}$ is, therefore, $zq
\times zq$. The question is to find inverse of the matrix 
${\Mat V}_{{\Mat g},{\Mat h}}^{(z)}$. 

It turns out that this inverse is in fact the matrix of the
same structure, namely
\begin{eqnarray}
\left(  {\Mat V}_{{\Mat g},{\Mat h}}^{(z)} \right)^{-1} = 
{\Mat V}_{{\Mat e},{\Mat f}}^{(z)} \ , && {\rm where}
\nonumber \\ && {\Mat e} = -  \left( {\Mat I} + (z-1) {\Mat h}
+ {\Mat g} \right)^{-1} {\Mat g}   \ \ {\rm and} \ \  {\Mat f}
= - \left( {\Mat I} + (z-1) {\Mat h} + {\Mat g} \right) ^{-1} 
{\Mat h} \ .
\end{eqnarray} The result can be easily proved using block
matrix multiplication rule.

\centerline{\bf Lemma 3.}

Consider an auxiliary problem of the scalar product
\begin{eqnarray}
\left< {\vec \rho}^{(qz)} \left| {\Mat W}^{(qz)} \right| {\vec
\rho}^{(qz)} \right> \ ,  \nonumber
\end{eqnarray} where ${\vec \rho}^{(qz)} = {\vec p}^{(q)}
\otimes {\vec i}^{(z)} = p_i$ (does not depend on replica
indices $\alpha$), and $ {\Mat W}^{(qz)} $ is block matrix
comprised of blocks $ {\Mat W}^{(q)}_{\alpha \beta}$.
Obviously, this scalar product is reduced to the scalar
products of smaller dimensionality $q$, that is, purely in
species space, summed over all the blocks of the matrix:
\begin{equation}
\left< {\vec \rho}^{(qz)} \left| {\Mat W}^{(qz)} \right| {\vec
\rho}^{(qz)} \right> = \left< {\vec p}^{(q)} \left|  \left(
\sum_{\alpha \beta} {\Mat W}^{(q)}_{\alpha \beta} \right)
\right| {\vec p}^{(q)} \right> \ .
\end{equation}

\newpage

\section{Relationship between the average number of
species-species contacts and the interaction matrix}

Note that this relation can be easily derived directly from
our formalism as well:   The Hamiltonian (\ref{eq:GenHam}) can
also be expressed directly in terms of the number of contacts
$n_{ij}$ between monomers of species $i$ and $j$:
${\cal H} =  \sum_{ij}^q B_{ij} n_{ij}$, where we have
previously substituted 
$n_{ij} = \sum_{I \neq J}^N \delta_{s_I,i} \delta_{s_J,j}
\delta({\Vec r}_I - {\Vec r}_J)$.  

Therefore, the average number of contacts can be directly
calculated in terms of the derivative of the free energy with
respect to
$B_{ij}$. However, at this point, we must indicate one point
in which we have been a bit cavalier in our previous
derivation.  Specifically, in order to perform the
Hubbard-Stratonovich transformation, we have summed over all
pairs of monomers $\sum_{I,J}$ instead of only the different
pairs
$\sum_{I \neq J}$.  This overcounting of self-site interaction
leads to a spurious term in the free energy ${\Mat
\Delta}{\Mat B}$.  Excluding this term from the free energy,
which is equivalent to performing the sum
$\sum_{I \neq J}$, carrying terms in the free energy to ${\cal
O}(B^2)$, and taking the derivative with respect to $B_{ij}$
yields
\begin{equation}
\left< n_{ij} \right> = p_i p_j \left(1 -  { B_{ij} \over T_m}
\right)
               \approx \  p_i p_j \exp \left( -  { B_{ij}
\over T_m} \right)
\end{equation}
of Ref.
\onlinecite{Gutin}, i.e. either $T_m = T_p$ for chains in the
target phase, $T_m = T_f$ for chains in the frozen phase, or
$T_m = T$ for chains in the random phase.

\begin{figure}
\caption{Phase diagram for different values of the matrix
similarity factor
$g$.  We find three phases for designed globular
heteropolymers: {\em Random}, in which many (${\cal O}(e^N)$)
conformations dominate equilibrium much as the equilibrium
conformation of a globular homopolymer;  {\em Frozen}, in
which only a few (${\cal O}(1)$) conformations dominate
equilibrium in a glass-like phase; and {\em Target} in which
only the target conformation (Native state) is found. For
decreasing values of $g$, the boundary of the target phase
moves to the left.  For example, consider the case in which one
performs a computer simulation of protein folding: nature has
``prepared'' proteins with the matrix of interactions ${\Mat
B}^p$  at some preparation temperature $T_p/T_{pf} < 1$ and one
wishes to renature these proteins with some simulated
potentials
${\Mat B}$ at some simulated acting temperature $1 < T/T_f <
T_{\rm tar}/T_f$; this desired pair of prepatation and acting
temperatures is signified by a  circle on the figure.  If one
could exactly reproduce the potentials used in Nature, i.e.
${\Mat B} = {\Mat B}^p$, then $g=1$ and the phase behavior is
unchanged from Naturally renatured proteins: the circle is
within the target phase.  If the potentials used for
renaturation are not precisely those used for preparation,
i.e. $g<1$, then these errors in the  potentials effect the
nature of the phases by    moving the  boundary of the target
phase to the left, thereby shrinking the target phase.  If the
errors are small compared with the degree of optimization,
renaturation to the target phase is still possible, i.e. as
shown in the figure, the circle is still in the target phase
for $g>0.95$.  Physically, the optimization of the native
state (``preparation'') allows some errors and this simply
leads to a less optimized native state.  Once these errors are
large enough to overcome the optimization, the system will no
longer renature to the target phase: in the figure we see that
for $g<0.95$, the circle is no longer in the target phase. }
\end{figure}


\begin{thebibliography}
\bibitem{Russian} This Russian sentence says: ``For example,
to the best of our hope, our reader considers this text to be
written in English and thus unlikely understands this
sentence''.
\end{thebibliography}
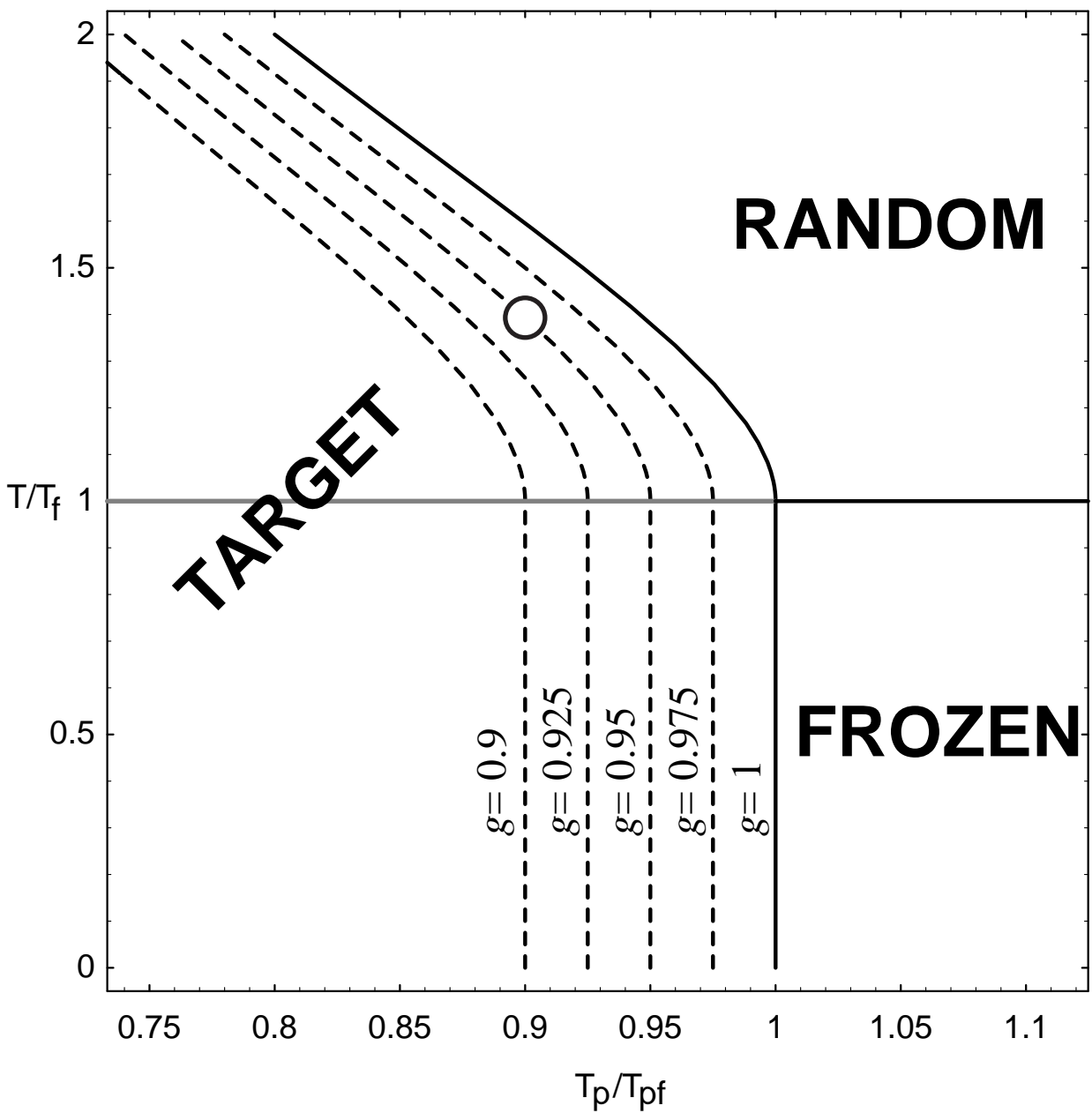
\end{document}